\documentclass[sigplan,screen]{acmart}

\usepackage[utf8]{inputenc} 
\usepackage[T1]{fontenc}    
\usepackage{hyperref}       
\usepackage{url}            
\usepackage{booktabs}       
\usepackage{amsfonts}       
\usepackage{nicefrac}       
\usepackage{microtype}      
\usepackage{comment}
\usepackage{listings}
\usepackage{xcolor}
\usepackage{graphicx}
\usepackage{wrapfig}
\usepackage{caption}
\usepackage{subcaption}
\usepackage{fancyvrb}
\usepackage[compact]{titlesec}
\usepackage{parskip}
\newcommand{\system}{ControlFlag}
\newcommand{\embf}[1]{\textbf{\emph{#1}}}
\newcommand{\h}{\hspace{0.1in}}

\definecolor{codegreen}{rgb}{0,0.6,0}
\definecolor{codegray}{rgb}{0.5,0.5,0.5}
\definecolor{codepurple}{rgb}{0.58,0,0.82}
\definecolor{backcolour}{rgb}{0.95,0.95,0.92}
\lstdefinestyle{mystyle}{
    keywordstyle=\color{black},
    basicstyle=\ttfamily\footnotesize,
    breakatwhitespace=false,         
    breaklines=true,                 
    captionpos=b,                    
    keepspaces=true,                 
    numbersep=5pt,                  
    showspaces=false,                
    showstringspaces=false,
    showtabs=false,                  
    tabsize=2
}
\title{ControlFlag: A Self-Supervised Idiosyncratic Pattern Detection System for Software  Control Structures}

\author{Niranjan Hasabnis}
\affiliation{\vspace{-1.5mm}\institution{Intel Labs, Santa Clara, CA, USA\vspace{-1.5mm}}
             \country{}}
\email{niranjan.hasabnis@intel.com}

\author{Justin Gottschlich}
\affiliation{\vspace{-1.5mm}\institution{Intel Labs, Santa Clara, CA, USA\vspace{-1.5mm}}
             \country{}}
\affiliation{\institution{University of Pennsylvania, Philadelphia, PA, USA\vspace{-1.5mm}}
             \country{}}
\email{justin.gottschlich@intel.com}

\setcopyright{acmlicensed}
\acmPrice{15.00}
\acmDOI{10.1145/3460945.3464954}
\acmYear{2021}
\copyrightyear{2021}
\acmSubmissionID{pldiws21mapsmain-p62-p}
\acmISBN{978-1-4503-8467-4/21/06}
\acmConference[MAPS '21]{Proceedings of the 5th ACM SIGPLAN International Symposium on Machine Programming}{June 21, 2021}{Virtual, Canada}
\acmBooktitle{Proceedings of the 5th ACM SIGPLAN International Symposium on Machine Programming (MAPS '21), June 21, 2021, Virtual, Canada}



\begin{document}
\begin{abstract}

Software debugging has been shown to utilize upwards of half of developers' time. Yet, \emph{machine programming} (MP), the field concerned with the automation of software (and hardware) development, has recently made strides in both research and production-quality automated debugging systems. In this paper we present \system, a \emph{self-supervised} MP system that aims to improve debugging by attempting to detect idiosyncratic pattern violations in software control structures. {\system} also suggests possible corrections in the event an anomalous pattern is detected. We present \system's design and provide an experimental evaluation and analysis of its efficacy in identifying potential programming errors in production-quality software. As a first concrete evidence towards improving software quality, {\system} has already found an anomaly in CURL that has been acknowledged and fixed by its developers. We also discuss future extensions of {\system}.

\end{abstract}
\maketitle

\section{Introduction}

According to some studies, upwards of 50\% of software development time is spent debugging~\cite{Britton:2012:reversibledebugging}. Therefore, even a minor reduction of debugging time could result in large software development savings, while simultaneously improving programmer productivity~\cite{gottschlich:pact:2012, gottschlich:2013:pact}. \emph{Machine programming} (MP), which is concerned with the automation of software (and hardware) development~\cite{gottschlich:2018:mapl}, has been shown as one technique to reduce software debugging time~\cite{alam:2019:neurips, dinella:2020:hoppity, goues:2019:cacm, gupta:2020:neurips}. Recently, there has been a flurry of research in MP due to advances in machine learning, formal methods, data availability, and computing efficiency, amongst other things~\cite{cen:2020:mapl, dinella:2020:hoppity, elboher:2020:cav, luan:2019:oopsla, odena:2020:iclr, ratner:2019:mlsys, Yasunaga:2020:ICML, ye:2020:misim}. In general, MP systems aim to improve programmer productivity and various quality characteristics of software, such as performance or security. Some examples of recent MP tools are automatic code generators~\cite{becker:2021:evosoft, kamil:2016:pldi, mandal:mlsys:2021}, assembly to IR translators~\cite{hasabnis:2016:eissec, hasabnis:2016:lisc}, code recommendation and similarity systems~\cite{luan:2019:oopsla, ye:2020:misim}, automated bug detection systems~\cite{dinella:2020:hoppity}, static and dynamic learned optimizations~\cite{cen:2020:mapl, patabandi:2021:maps}, performance regression test generation~\cite{alam:2019:neurips}, and automatic completion of program constructs for integrated development environments (IDEs)~\cite{gao:2020:OOPSLA, Svyatkovskiy:2019:KDD}.

In this paper, we present an MP approach to automatically identify violations of programming patterns. Violations of programming patterns can be thought of as syntactically-valid code snippets that deviate from a typical usage of the underlying code constructs (i.e., programming pattern). Consider the following example of a possible violation of a common use of a \texttt{for} loop in the C programming language:

\begin{footnotesize}
\centering
\begin{Verbatim}[commandchars=\\\{\}]
1. \textcolor{blue}{for} (\textcolor{blue}{int} i = 0; i < N; i++) \{
2.    // Some program code
3.    i++;
4. \}
\end{Verbatim}
\end{footnotesize}

This example is a possible violation of the typical usage of a \texttt{for} loop because it increments the loop counter twice in a single loop pass (lines 1 and 3). The typical use of a loop counter is to increment it once per pass. 
Violations can be found in other languages as well. Consider the following violation example in Verilog~\cite{sutherland:2006:SNUG}:

\newpage
\begin{footnotesize}
\centering
\begin{Verbatim}[commandchars=\\\{\}]
1.  \textcolor{blue}{reg} [1:0] state;
2.  
3.  \textcolor{blue}{always} @(state)
4.  \textcolor{blue}{case} (state)
5.    00: // do State 0 stuff
6.    01: // do State 1 stuff
7.    10: // do State 2 stuff
8.    11: // do State 3 stuff
9.  \textcolor{blue}{endcase}
\end{Verbatim}
\end{footnotesize}

In this example, case $10$ and $11$ (lines 7 and 8) are unreachable as \texttt{state} is a binary variable (2-bits), but the default numerical base in Verilog is decimal. Note, however, that these examples \emph{may} or \emph{may not} be bugs. It could be the case that the programmer intended to increment the loop counter twice in the first example and meant to combine binary variables with decimal numbers in the second example. In that regard, we argue that the problem of identifying idiosyncratic programming violations is \emph{probabilistic} in nature. 

In this work, we present \system, a \emph{self-supervised} system that automatically identifies potential errors in C/C++ \texttt{if} statements, one of the core control structures of the C family of languages. In addition to identifying such potential errors, \system\ produces suggestions as corrections for the potential errors it finds. Other examples of control structures in high-level languages such as C and C++ are: \emph{(i)} selection statement: \texttt{if}, \texttt{else if}, \texttt{else} and \texttt{switch}, \emph{(ii)} repetition statements: \texttt{for} loops, \texttt{while} loops, \texttt{do} \texttt{while} loops, \emph{(iii)} jump statements: \texttt{goto}, \texttt{throw} statements. Consider the following C++ code:

\begin{footnotesize}
\centering
\begin{Verbatim}[commandchars=\\\{\}]
                    \textcolor{blue}{if} (x = 7) y = x;
\end{Verbatim}
\end{footnotesize}

In the above code, it is likely that the programmer's intention is to assign \texttt{x}'s value to \texttt{y} only when \texttt{x} is equal to \texttt{7}. Unfortunately, due to the omission of the second \texttt{=} operator in the \texttt{if} conditional expression, the intended equality check is transformed into an assignment operation. This results in \texttt{x} being assigned the value of \texttt{7}. Because \texttt{x}'s value is non-zero, the \texttt{if} condition always returns true. The resulting true condition then causes \texttt{y} to always be assigned the value stored in \texttt{x} (which, coincidentally, is always $7$ due to the \texttt{if}'s assignment). The code to check \texttt{x}'s equality to $7$ is:

\begin{footnotesize}
\centering
\begin{Verbatim}[commandchars=\\\{\}]
                    \textcolor{blue}{if} (x == 7) y = x;
\end{Verbatim}
\end{footnotesize}

As was the case in the prior examples, we can only speculate that the original code did not properly capture the programmer's intention. Consequently, and more generally, any potential recommendation by \system\ is probabilistic. In fact, it is from the analysis of various repositories of code that \system\ makes a determination about whether a particular control structure is a potential error, based on a recurrence of commonality of idiosyncratic patterns found in the code repositories it analyzes.

Identifying typographical coding errors such as these can be challenging for at least two reasons. First, the assignment of variables within conditionals is legal syntax in C/C++ (e.g., \texttt{if (x = 7)}). As such, this type of intent-error might not be flagged by a compiler because the code's syntax is, in fact, legal. Second, compilers and static analyzers can use data-flow analysis to identify such errors\footnote{An optimizing compiler (e.g., GCC -O2) can eliminate \texttt{if} statement and replace it by \texttt{x = 7; y = 7;}, eliding away the \texttt{if} condition entirely.}, but data-flow analyses have their own limitations (e.g., locally scoped versus globally scoped analyses, etc.). Nonetheless, compilers such as GCC and LLVM already use a rules-based approach to warn programmers in a variety of potentially erroneous cases. For instance, GCC-10.2.0's \texttt{-Wall} option warns programmers of the above code as:

\begin{footnotesize}
\begin{center}
\begin{Verbatim}[commandchars=\\\{\}]
  test.cpp:3:9: \textcolor{red}{warning}: suggest parentheses around
    assignment used as truth value [-Wparentheses]
                    \textcolor{blue}{if} (x = 7) y = x;
                        ~~^~~
\end{Verbatim}
\end{center}
\end{footnotesize}

However, rules-based approaches tend to have at least two core limitations. First, they can be labor-intensive. New rules generally need to be added to flag new types of potential errors, especially ones created by evolving programming languages constructs~\cite{dinella:2020:hoppity}. Second, many advanced warning systems may require a compilable program. For example, compiler-based warnings -- like the GCC warning above -- requires code that is compilable to flag such issues. Moreover, it may be unlikely that such compiler-based approaches are practical in live development environments where a recommendation system may attempt to dynamically identify issues as the programmer is writing code (where such code is often incomplete and not compilable).

In this paper, we take a statistical approach to identifying programming pattern violations by recasting them as anomalies. Likewise, our ControlFlag system is designed to behave as an anomalous code detector. We hypothesize that \emph{using certain patterns (such as assignment) inside an} \texttt{if} \emph{statement in C/C++ language is relatively rare}. We test this hypothesis by mining idiosyncratic patterns found in the control structures of C/C++ programs found in open-source GitHub repositories. The mined patterns form a \emph{dictionary} that is then used to check a user's typed pattern and suggest automatic corrections in case of divergence. An advantage of such an approach is that it does not require labeled training data. {\system} uses the mined patterns from \emph{semi-trusted} GitHub repositories in a \emph{self-supervised} fashion, eliminating the need for labeled code anomalies.

In addition to the examples discussed earlier, {\system} can learn many other types of idiosyncratic patterns. One pattern category is in the space of programming language typing rules and their binding to proper mathematical operators. These rules can be used to flag anomalies related to the uses (and misuses) of types. For instance, we used \system\ to detect an integer and Boolean type mismatch in the CURL open-source project, which led to the CURL development team to redesign a small portion of their code to correct the issue (details forthcoming). Another pattern category is around memory management programming patterns. \system\ learned that it is often appropriate to ensure pointers are evaluated against \texttt{NULL} before being dereferenced and will flag missing \texttt{NULL} pointer checks.

This paper makes the following technical contributions:
\begin{itemize}
    \item We present {\system}, which, to our knowledge, is the first-of-its-kind \emph{self-supervised} idiosyncratic programming pattern detection system. 
    
    
    \item While we only demonstrate {\system} for C/C++, we have designed it to be programming language agnostic. As such, it should be capable of learning idiosyncratic signatures of any type of control structure in any programming language.
    
    \item We present initial results of \system's ability to identify idiosyncratic pattern violations in \texttt{if} statements of C/C++ programs. These results span $\approx{6,000}$ GitHub repositories and over one billion lines of code. 

    \item We provide a concrete illustration of \system's capabilities on the client URL (CURL) open source project, where it was able to identify a code anomaly that their development team was unaware of. The CURL team agreed with \system's findings and has since upstreamed a fix to address the issue.
    
    
\end{itemize}

\section{\system{} Design}
\label{section:design}

\begin{figure*}[htpb]
\begin{center}
\includegraphics[width=0.9\textwidth]{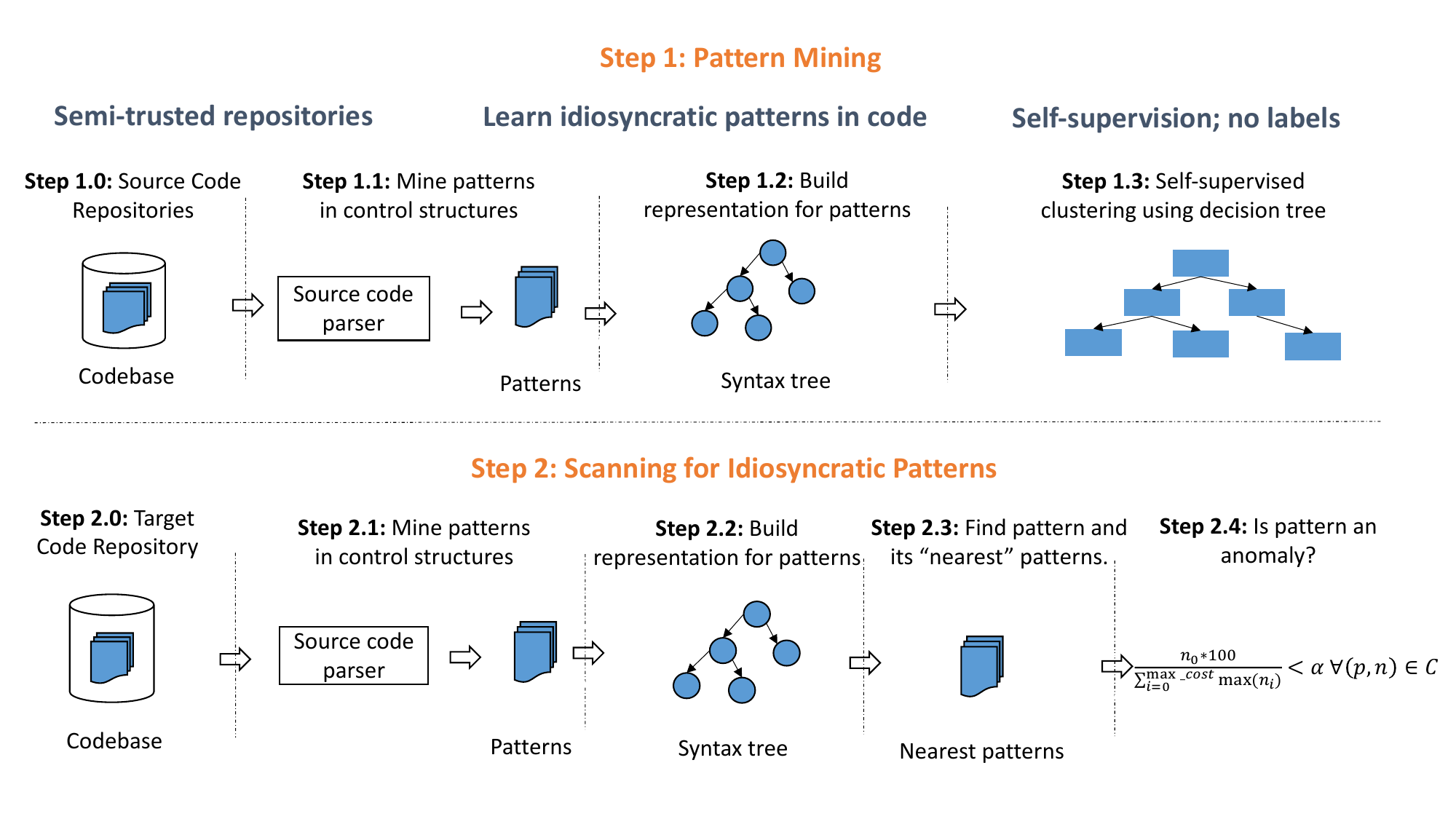}
\vspace{-0.15in}
\caption{Overview of \system{} System.}
\label{fig:system_overview}
\end{center}
\end{figure*}

Figure~\ref{fig:system_overview} provides an overview of {\system}, consisting of two main phases: \emph{(i)} pattern mining and \emph{(ii)} scanning. The \emph{pattern mining} phase consists of learning the common (and uncommon) idiosyncratic coding patterns found in the user-specified GitHub repositories, which, when complete, generates a precedence dictionary that contains acceptable and unacceptable idiosyncratic patterns. The \emph{scanning} phase consists of analyzing a given source code repository against the learned patterns for possible anomalies. When anomalous patterns are identified, {\system} flags them and provides alternative coding recommendations.

\subsection{Pattern Mining}

The pattern mining phase is further divided into the following sub-phases: \emph{(i)} source repository selection, \emph{(ii)} code parsing, and \emph{(iii)} decision tree construction.

\paragraph{Source repository selection.}
In our experiments, we mined patterns from GitHub repositories that met certain minimum number of stars. In other words, we used the number of stars of a GitHub repository as a measure of its quality. Without any easy and quantifiable measure of the quality of the programs in those repositories other than GitHub stars, we believe that such a selection gives us semi-trust in the data. We, however, believe that this is an open problem as quality is a subjective term.

\paragraph{Parsing control structures.}
After a source repository is selected, {\system} generates a list of programs in the high-level programming language of interest. Every program from this list is then parsed. The resulting parse trees -- currently in the form of abstract syntax trees (ASTs) -- are then converted into a higher-level abstracted tree to prune unnecessary details. While {\system} parses these programs, it does not discard programs containing parse errors. For {\system}'s purpose, all that is needed is that control structures do not have parse errors; parse errors elsewhere in the code may not impact \system's functionality. We believe this characteristic of {\system} is important, given that we have found that the majority of the C/C++ programs found in open-source are not compilable. Additionally, such an approach would enable {\system} to eventually be used in a live programming environment or an integrated development environment (IDE).

\paragraph{Levels of abstractions.}
We used TreeSitter APIs~\cite{treesitter} to parse the control structures and used its AST representation\footnote{To keep the description brief, we do not specify its grammar here.} as the lowest-level tree representation (referred to as an L1 abstraction level). But during the experimentation, we found that AST representation would capture exact structural form of the expressions inside the control structures. For instance, for expression from C/C++ language such as \texttt{a.b.c \& (1 << e))}, the AST would be:

\begin{footnotesize}
\begin{center}
\begin{verbatim}
(binary_expr ("&") 
  (field_expr
    (field_expr ((identifier)(field_identifier)))
    (field_identifier))
  (parenthesized_expr
    (binary_expr ("<<")(number)(identifier))))
\end{verbatim}
\end{center}
\end{footnotesize}

We observed that capturing such a precise structural form would later reduce the chances of matching a target pattern during the scanning phase. The higher-level abstraction level (referred to as an L2\footnote{We use L1 and L2 terminology from the concept of caches in computer architecture. An item missing at L1 has a chance of being present at L2.} abstraction level), developed on top of an AST, drops the precision in the structural form by keeping the ASTs to a finite and small height. For that purpose, it introduces a new tree node type (named \texttt{non\_terminal}) to represent pruned subtrees. The higher-level tree for the aforementioned AST would then take the form:

\begin{footnotesize}
\begin{center}
\begin{verbatim}
(binary_expr ("&") (non_terminal_expr)(non_terminal_expr))
\end{verbatim}
\end{center}
\end{footnotesize}

Note that the higher-level tree above drops both the children of \texttt{\&} and marks them as \texttt{non\_terminal}. This would intuitively increase the number of false negatives. Nonetheless, it allows \system\ to check for idiosyncratic violations in \texttt{\&} during the scanning phase rather than declaring the whole pattern as absent from the dictionary.

\paragraph{Decision tree construction.}
During parsing, a pattern at both the L1 and L2 abstraction levels can be dumped out in textual format using a tree traversal. In fact, we represent the mined patterns internally by building a prefix tree\footnote{In our experiments, we found that prefix tree structure performed reasonably well (time complexity of search and space usage) for our experiments. We have not yet invested in optimizing it to reduce the space usage, but such an optimization using deterministic finite automaton (DFA) seems possible.} (or trie~\cite{cormen:2009:algo}) over the text strings for the trees of the mined patterns. Every path in the prefix tree that ends with a terminal node, corresponding to a valid pattern, also stores the number of occurrences of that pattern. We build different prefix trees for text strings that are at different levels of abstractions. By doing this, we can convert idiosyncratic patterns into different abstraction levels and check them against their respective prefix trees. 

\subsection{Scanning for Unusual Patterns}

When a user specifies a target repository to scan for unusual patterns, {\system} first obtains a list of idiosyncratic patterns that occur in the specified control structures. For every pattern in the list, it builds a parse tree at an L1 abstraction level and checks it against the prefix tree at L1 level. If the pattern is present in the tree, it skips the check against the prefix tree at L2 level. If the pattern is missing, \system\ checks it against the prefix tree at L2 level. {\system} triggers an automatic correction phase irrespective of whether a pattern is found or not found (at L1 and L2 levels). In the former case, although the pattern was found, it could be rare; in the latter case, it is known to be rare, because it is absent. \system's automatic correction phase then flags the pattern if it is anomalous based on a heuristically calculated threshold. If the pattern is flagged as anomalous, \system\ suggests possible corrections based on tree similarity criteria.

\paragraph{Automatic correction.}
Automatic correction of strings~\cite{cormen:2009:algo} is a well-studied problem in computer science. In our first embodiment of {\system}, we use edit distance between strings (using dynamic programming algorithm) to suggest possible corrections of the target string. 

We experimented with three approaches to suggest corrections of a possibly-erroneous target pattern: a naive approach, Norvig et al.~\cite{norvig_autocorrection}, and Symmetric Delete~\cite{symmetric_delete}\footnote{As the details of these approaches are not necessary to understand the core concepts of the main paper, we place a more detailed analysis of these techniques in Appendix~\ref{section:appendix_autocorrection}.}. In our experiments, we found that the temporal and spatial complexity of the naive approach is reasonable compared to Norvig et al.'s approach and symmetric delete, both of which encountered super-linear temporal and spatial complexity. Consequently, we used naive approach for our evaluation.
We have, nonetheless, optimized it in the following obvious ways: \emph{(i)} caching the results of automatic correction process, \emph{(ii)} compacting the string representation of the parse trees by using short IDs for unique tree nodes, and \emph{(iii)} employing parallelism while traversing the prefix trees.

\paragraph{Ranking the results of automatic correction.}
The outcome of the automatic correction process is a list of possible corrections, where every correction contains its occurrences in the dictionary and its edit distances from a given target string. When {\system} presents the auto-correction results to the user, it first sorts them in the increasing order of an edit distance and then by the number of occurrences. This simple heuristic is based on the intuition that the typographical violation of one character has a greater probability than a typographical violation in more than one character. \system\ uses the sorted results of the automatic correction process to determine a threshold level to flag anomalies.

\paragraph{Anomaly threshold.}
\system\ uses two criteria for the purpose of declaring a certain pattern as anomaly.
First, a target pattern that is missing from the dictionary is declared anomalous because it is missing. Second, a target pattern that is not missing, but has its automatic correction results satisfy the following formula, is declared anomalous.
    $$\frac{n_0 \times 100}{\sum_{i=0}^{max\_cost} max(n_i)} < \alpha \h\forall(p,n)\in C$$
$C$ is the set of automatic correction results, in which every result contains a corrected pattern $p$ and its occurrences $n$. $\alpha$ is a user-defined anomaly threshold, and $max$ is a function that calculates the maximum of a list of occurrences.

Intuitively, we calculate the percentage contribution of the number of occurrences of a possibly-incorrect target pattern ($n_0$) against the maximum number of occurrences at every edit distance. In other words, if the possible corrections of a target string have high frequency than the target string at smaller edit distances (such as 1), then it is likely that the target string is a violation. $\alpha$ is the user-controllable anomaly threshold that is set to 5\% by default. 

\section{Experimental Evaluation}
\label{section:evaluation}

In this section, we present results of {\system} in identifying idiosyncratic pattern violations in \texttt{if} statements of C/C++ programs.

\subsection{Setup}

All the experiments were performed on a 56-core Intel Xeon Platinum 8280 CPU, using $\approx{200}$GB of memory and hyper-threading enabled. The server was running CentOS-7.6.1810 operating system and GCC-10.2 compiler.

\textbf{Source repository selection.}
For the pattern mining phase, we chose the top 6000 open-source GitHub repositories that used C/C++ as their primary programming language and had received at least 100 stars. As previously mentioned, {\system} uses GitHub stars as a mechanism to infer quality (i.e., \emph{semi-trust}) of the source repositories used for training.

\textbf{Target repository selection.}
In our experiments, we used following open-source popular projects to scan for violations of typical programming patterns: \texttt{OpenSSL-1.1.1h}, \texttt{CURL-7.73}, \texttt{FFmpeg-n4.3.1}, \texttt{git-2.30}, \texttt{vlc-4.0}, \texttt{lxc-4.0}, \texttt{lz4-1.9.3}, and \texttt{reactos-0.4.13}. There was no particular reason to choose these packages besides the fact that they are widely used open-source software packages and are principally implemented using C/C++ programming language.

\subsection{Results}

\begin{figure}[!t]
\includegraphics[width=0.45\textwidth]{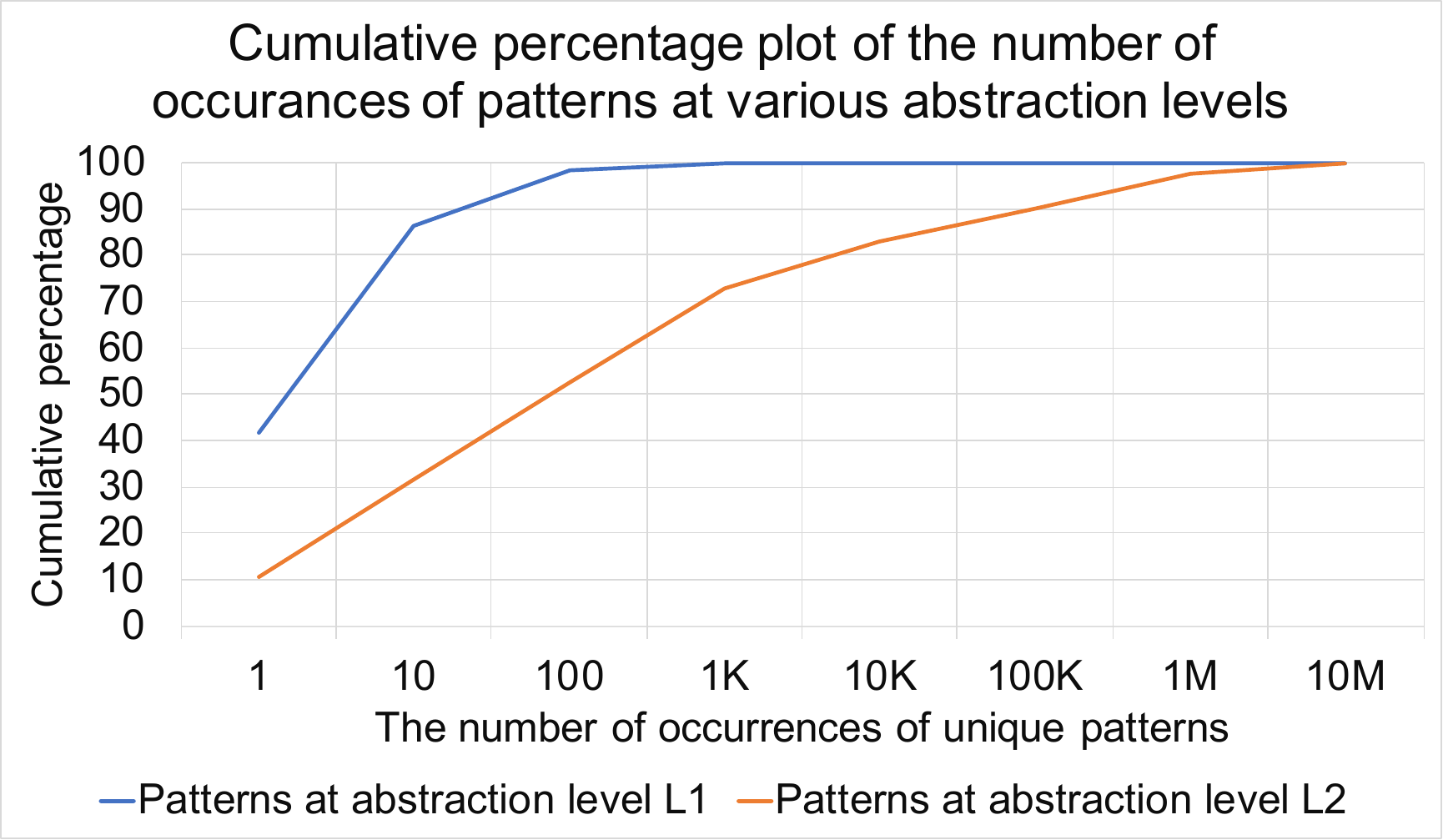}
\caption{Cumulative percentage plot of the number of occurrences of unique patterns at different abstraction levels.}
\label{fig:cpp_source_patterns}
\end{figure}

\textbf{Mining patterns from source repositories.}
We used 6000 repositories for the pattern mining phase, and they consisted of a total of 2.57M C/C++ programs. These programs had $\approx{38}$M total patterns, $\approx{831}$K unique patterns at the L1 abstraction level, and 468 unique patterns at the L2 abstraction level. Figure~\ref{fig:cpp_source_patterns} shows the cumulative percentage plot of the number of occurrences of unique patterns at both the abstraction levels. As expected, $\approx{90}$\% of the unique patterns at the L1 level have low frequency (close to 10). At L2 level, however, $\approx{90}$\% of the patterns have higher frequency because of the grouping of multiple patterns at L1 level.

Table~\ref{table:most_frequent_source_patterns} shows top-10 frequently occurring patterns at the L1 abstraction level from the training dataset. Most of these patterns are expected; it is also good to see the NULL check in the list --- in our opinion, it talks about good programming practice. Table~\ref{table:least_frequent_source_patterns}, on the other hand, shows patterns, such as \texttt{if (x = 7)}, that have less than 1\% occurrences of 38M. It is interesting to see bitwise operators in C/C++ (such as \texttt{|} and \texttt{$^\wedge$}) in the list. We believe that these operators are more common in low-level code that operates close to hardware. This observation also suggests that the selection strategy for source repositories could be different in which we consciously ensure a uniform mix of repositories that contain code belonging to different technical domains.

\begin{table}
\caption{Some of the most frequently occurring patterns at L1 abstraction level}
\label{table:most_frequent_source_patterns}
\vspace{-0.1in}
\begin{footnotesize}
\begin{tabular}{l|r|l}
\hline
AST & Occur- & Example C \\
 & rences & expressions \\ \hline \hline
\texttt{(id)} & 4.3M & \texttt{if (x)} \\
\texttt{(unary\_expr (``!'') (id))} & 2.09M & \texttt{if (!x)} \\
\texttt{(field\_expr (id)(field\_id))} & 1.3M & \texttt{if (p->f)} \\
\texttt{(binary\_expr (``=='') (id)(id))} & 1.16M & \texttt{if (x == y)}\\
\texttt{(binary\_expr (``<'') (id)(number))} & 1.13M & \texttt{if (x < 0)}\\
\texttt{(binary\_expr (``=='') (id)(number))} & 1.09M & \texttt{if (x == 0)}\\
\texttt{(call\_expr (id)(arg\_list (id)))} & 1.05M & \texttt{if (foo(x))}\\
\texttt{(binary\_expr (``=='') (id)(null))} & 790K & \texttt{if (p == NULL)} \\
\texttt{(binary\_expr (``=='') (field\_expr} & & \\
\h\h\texttt{(id)(field\_id))(id))} & 732K & \texttt{if (p->f == y)}\\
\texttt{(binary\_expr (``!='') (id)(id))} & 636K & \texttt{if (x != y)}\\
\hline
\end{tabular}
\end{footnotesize}
\end{table}

\begin{table}
\caption{Patterns with less than 1\% occurrences at L1 abstraction level}
\label{table:least_frequent_source_patterns}
\vspace{-0.1in}
\begin{footnotesize}
\begin{tabular}{l|r|l}
\hline
AST & Occur- & Example C \\
 & rences & expressions \\ \hline \hline
\texttt{(binary\_expr (``='') (id)(number)))} & 487 & \texttt{if (x = 0)} \\
\texttt{(binary\_expr (``='') (id)(id)))} & 476 & \texttt{if (x = y)} \\
\texttt{(binary\_expr (``='') (id)(call\_expr} & & \\
\h\h\texttt{(id)(arg\_list))))} & 356 & \texttt{if (x = foo(y))} \\
\texttt{(binary\_expr (``\%'') (id)(number)))} & 6468 & \texttt{if (x \% 2)} \\
\texttt{(binary\_expr (``|'') (id)(id)))} & 1137 & \texttt{if (x | y)} \\
\texttt{(binary\_expr (``$^\wedge$'') (id)(id)))} & 813 & \texttt{if (x $^\wedge$ y)} \\
\texttt{(binary\_expr (``=='') (number)} & &  \\
\h\h\texttt{(number)))} & 236 & \texttt{if (0 == 0)} \\
\hline
\end{tabular}
\end{footnotesize}
\end{table}

Mining patterns from 6000 source repositories with 56 worker threads took approximately two hours. Building the prefix trees at the L1 and L2 abstraction levels from those patterns took approximately three minutes. We also dumped the patterns at the L1 and L2 levels in a textual format into a file, which was $\approx{11}$GB in size. The memory consumption of {\system} after building the prefix trees was $\approx{8}$GB, which has a reduced spatial footprint due to compression performed by the prefix tree.

\begin{figure*}[htbp]
\begin{center}
\begin{subfigure}[b]{0.45\textwidth}
\centering
\includegraphics[width=\textwidth]{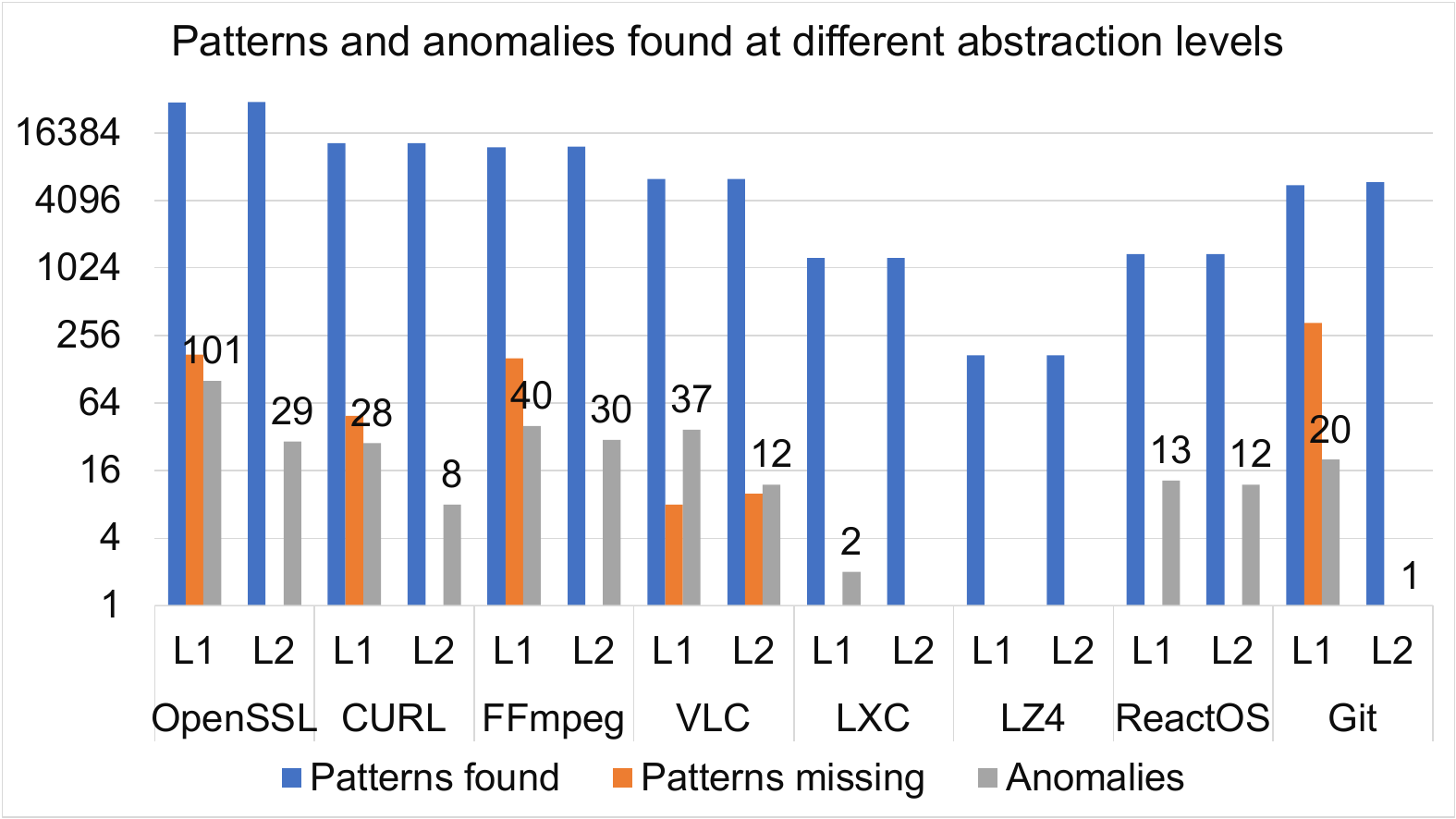}
\caption{Patterns and anomalies at different abstraction levels}
\label{fig:eval_patterns_found_missing}
\end{subfigure}
\hfill
\begin{subfigure}[b]{0.45\textwidth}
\centering
\includegraphics[width=\textwidth]{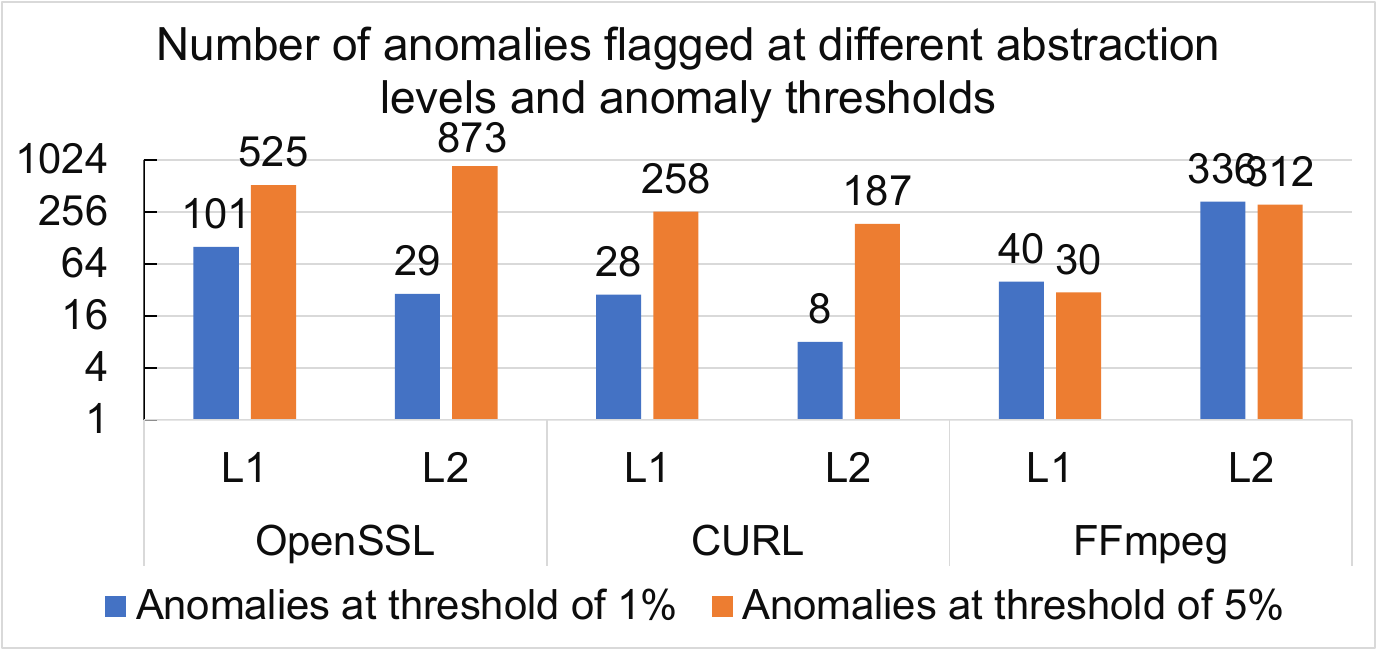}
\caption{Effect of different thresholds on the number of anomalies}
\label{fig:eval_patterns_flagged_anomalies}
\end{subfigure}
\end{center}
\vspace{-0.1in}
\caption{Scanning popular software packages for typographical errors}
\label{fig:evaluation_results}
\end{figure*}

\textbf{Scanning target repositories.}
After pattern mining phase, \system\ obtained C/C++ programs from the packages from the evaluation set and scanned them for violations of programming patterns in \texttt{if} statements. The packages contained on the order of a few thousand files, with \texttt{FFmpeg} containing the most (3670) and \texttt{lz4} containing the least (51). With 56 scanner threads, {\system} took $\approx{4}$ hours to scan FFmpeg and 10 minutes to scan lz4. 

Figure~\ref{fig:eval_patterns_found_missing} and Figure~\ref{fig:eval_patterns_flagged_anomalies} show the results of the scanning phase. Figure~\ref{fig:eval_patterns_found_missing} shows the effect of two abstraction levels on the number of patterns that are found and missing in the prefix trees. The figure also shows the number of flagged anomalies at different abstraction levels. In summary, all the patterns are found at the L2 level (except for \texttt{VLC}, which had 10 missing at L2 level), while a few are missing at the L1 level. Figure~\ref{fig:eval_patterns_flagged_anomalies} shows the effect of different anomaly thresholds (1\% and 5\%) on number of patterns flagged as anomalies. As expected, the number of anomalies flagged at 5\% are higher than that at 1\%. This raises an obvious and deeply studied question: what should the anomalous threshold be set to? We could set the threshold to a value smaller than 1\%, given that Figure~\ref{fig:cpp_source_patterns} shows that many patterns have small frequencies at the L1 level. Yet, because the absolute number of anomalies flagged at the L1 level (after removing duplicates) are reasonable for a manual inspection, we chose not to experiment with reducing the anomaly threshold further.

\textbf{Anomalies flagged in the scans.}
We now discuss some of the reported anomalies in our experiments. As OpenSSL had the highest number of anomalies, we chose OpenSSL as the candidate for analysis. We also discuss an anomaly from CURL that led to software change. A more detailed report of these anomalies is provided in Appendix~\ref{section:appendix_result}. Note that we have not yet confirmed if these anomalies are bugs --- we can only confirm that by applying the changes suggested in the automatic corrections and running the validation tests (or contacting the developers).

\textbf{\textit{Anomaly 1.}} CURL's \texttt{lib/http\_proxy.c} uses \texttt{s->keepon > TRUE} expression at line number 359. This was flagged as anomalous because the training dataset had only 4 patterns that contained a boolean value \texttt{TRUE} in \texttt{>}. \system's top suggested correction \texttt{s->keepon > number}, at edit distance of 2, had 127K occurrences. We found that \texttt{s->keepon} is of type \texttt{int} in CURL, while \texttt{TRUE} is defined as \texttt{true}, which is defined as integer \texttt{1} in C language. So this expression is a comparison between a signed 32-bit integer and an integer value of a boolean, which is why GCC did not flag it. We believe that \texttt{> true} expression, however, is ambiguous for two reasons: boolean values are typically used with logical and bitwise operators, and, in C language, any non-zero value is considered as \texttt{true}. We conveyed this ambiguity to CURL developers~\cite{curl_discussion} and proposed \texttt{s->keepon > 1} as a better expression. They acknowledged the ambiguity and resolved it~\cite{curl_fix} by using \texttt{enum} type to model different values of \texttt{keepon}. We believe this example demonstrates early promise of \system\ in that it found an anomaly, which has since improved the robustness of CURL by eliminating the potential ambiguity.

\textbf{\textit{Anomaly 2.}} OpenSSL's \texttt{test/testutil/tests.c} uses expression \texttt{(s1 == NULL) $^\wedge$ (s2 == NULL)} at line number 268. The expression was flagged as anomalous because it had only 8 occurrences in the training dataset. While the top two suggested corrections --- \texttt{(s1 == NULL) | (s2 == NULL)} at edit distance 1 and \texttt{(s1 == NULL) || (s2 == NULL)} at edit distance 2 --- had 32 and 6808 occurrences in the training dataset, respectively.

On a similar note, we also found that OpenSSL uses expression \texttt{(m1 == NULL) == (m2 == NULL)}. This expression was flagged as anomalous because it had only 7 occurrences in the dictionary, while its possible corrections --- \texttt{(m1 == NULL) != (m2 == NULL)} at edit distance 1 and \texttt{(m1 == NULL) || (m2 == NULL)} at edit distance 2 --- had 27 and 6808 occurrences, respectively.

\textbf{\textit{Anomaly 3.}} OpenSSL's \texttt{test/evp\_test.c} uses expression \texttt{(-2 == rv)} at line number 1898. It was flagged as anomalous because it had 16529 occurrences in the training dataset, while its possible correction \texttt{variable == rv} at edit distance 1 had 1.1M occurrences. We believe that the base expression has lower occurrences because fewer programmers use that style of writing an equality check. We, however, believe that \texttt{number == variable} is a better style than \texttt{variable == number}, as it avoids possible typographic errors because compiler's lvalue check will prevent assignment to a constant.

\textbf{\textit{Anomaly 4.}} OpenSSL's \texttt{crypto/x509/t\_req.c} uses expression \texttt{(BIO\_puts(bp, ":") <= 0)} at line 141. This expression was flagged as anomalous as it had 475 occurrences in the training dataset. What we find interesting about this expression is that it compares the result of a function call with 0 and negative values, which is OpenSSL's approach for evaluating error codes. \system's top two suggested corrections were \texttt{(BIO\_puts(bp, ":") == 0)} and \texttt{(BIO\_puts(bp, ":") < 0)}, which, based on the data we analyzed, seemed to indicate more appropriate patterns --- 0 being a successful return code (as in standard libc) and comparison with the negative values for erroneous return codes. OpenSSL's expression somehow combines both the typical patterns together, resulting in a highly abnormal combination.

\subsection{Results Analysis and Future Directions}

Our analysis revealed that several of the reported anomalies looked innocuous and can be considered as false positives. We, however, did not tag them as false positives because, being an unsupervised approach, we do not have a list of anomalous patterns. Moreover, it seems that in such case we can confirm those anomalies by applying the correction suggested by {\system} and running sanity tests. On the other hand, we can possibly compile a list of known anomalies (as a ground truth) to measure precision and recall of our approach. For the current version, we kept this as future work. The reported anomalies, nonetheless, point to some interesting observations that guide us to fine-tune and extend {\system}. We discuss them below.

\paragraph{Repository-specificity.}
Some of the anomalies flagged in our experiments appear to be repository-specific coding styles for conditional expressions. A potential way to remove these anomalies could be to build a repository-specific style dictionary. Such an approach, however, may lead to an increase in the number of false negatives. A second alternative could be to build a generic dictionary from multiple repositories that is fine-tuned afterwards for the specific target repository.

\paragraph{Source repository selection.} We believe that GitHub stars is an indirect measure of quality of a source repository. It would be interesting to explore applicability of data preprocessing techniques (e.g., data cleansing) to, first of all, define quality and then filter out low-quality repositories.
 
\paragraph{Nested expressions.} Another important feature that {\system} could have is breaking down nested expressions into fix-sized expressions. 
Given that expression trees could be of infinite depth in most high-level languages, this feature could reduce the number of false positives. L2 abstraction level mentioned earlier (Section~\ref{section:design}) already does this to some extent. Nonetheless, it seems that a much-refined scheme could be conceived. Specifically, it would be interesting to find out the minimum height of the expression trees that would still preserve possible relationships between multiple subtrees of an expression tree. For instance, \texttt{p != NULL \&\& p->f} is a typical expression used in C/C++ to ensure that a pointer is non-NULL before a dereference. Breaking down this expression into trees of height 1 would lead to missing the ordering relationship between \texttt{p != NULL} and \texttt{p->f} --- \texttt{p != NULL} must occur before \texttt{p->f}.


\section{Related Work}

In this section, we discuss existing work in machine programming that is relevant to this paper.

\paragraph{Mining code patterns/idioms.} Given the vast amount of publicly-accessible open-source code, a number of research efforts have been devoted to extract useful information from code. For instance, mining syntactic fragments (also called as idioms) has been an active area of research~\cite{allamanis:2014:fse, allamanis:2018:icse, di:2019:SATToSE, jacob:2010:se, lozano:2010:WCRE, orlov:2020:LAFM, pham:2019:ICDS}. Code idioms, e.g., nested loops, exception handlers, etc., are used by source-code editors (IDEs) to assist programmers in writing code. Although IDEs typically enable programmers to manually add idioms to their IDEs, programmers may not be familiar with the latest idioms. To address this problem, techniques from data mining, frequent tree mining~\cite{zaki:2005:ieeecs}, and probabilistic grammars~\cite{allamanis:2014:fse} can be applied to mine idioms from source code repositories. While code idioms can, conceptually, be thought of as patterns, they differ from patterns in a sense that idioms represent ``interesting'' and ``useful'' code fragments that assist programmers in writing code. Our focus in this research is on detecting violations of programming patterns and not on assisting programmers in writing code. In other words, frequent patterns are ``interesting'' patterns in our case.

Mined code idioms can be used to solve other problem also. For instance, for the problem of \emph{program synthesis and semantic parsing}, --- where the goal is to generate a high-level language program that implements often-incomplete program specification, --- code idioms can capture semantic concepts that can simplify program synthesis~\cite{Iyer:2019:arxiv:LearningPI, shin:2019:arxiv}. Specifically, PATOIS system~\cite{shin:2019:arxiv} trains a program synthesizer to use code idioms. For the problem of semantic parsing --- where the program specification is in a natural language --- the research effort~\cite{Iyer:2019:arxiv:LearningPI} proposes an iterative method to extract code idioms and train semantic parsers using them. Frequent code patterns have also been used to \emph{learn typical API usages}~\cite{Fowkes:2016:FSE, Xie:2006:MSR} and \emph{detect API misuses}~\cite{acharya:2009:FASE, nielebock:2020:arxiv}. If learned API usage patterns (e.g., sequence of calls, method arguments and their order, etc.) are considered good/correct examples, then API misuses can be considered anomalies. The problem of detecting API misuses can then be considered conceptually similar to the problem of detecting abnormal programming patterns in {\system}. However, it could also be considered as a subset of the problem approached by {\system}, when API usage patterns are represented using ASTs. Several other representations of API usage patterns, such as call pairs~\cite{weimer:2005:springer}, association rules~\cite{livshits:2005:SE}, call sequences~\cite{thummalapenta:2007:ASE}, trees~\cite{allamanis:2014:fse}, and graphs~\cite{nguyen:2009:FSE}, exist nonetheless.

\paragraph{Automated bug detection, software defect prediction, and program repair.} As {\system} suggests automatic corrections for anomalous patterns, it can be considered to be close to the problem of \emph{automated bug detection and program repair}. Automated bug detection and program repair are growing and active areas of research in MP~\cite{alam:2019:neurips, allamanis:2018:learning, dinella:2020:hoppity, pradel:2018:OOPSLA, vasic:2019:iclr}. Most of these techniques rely on a learning based approach to detect and fix bugs. Hoppity~\cite{dinella:2020:hoppity}, in particular, uses deep learning model to detect and correct bugs in JavaScript code.

\emph{Software defect prediction}, a problem that is conceptually close to bug detection, attempts to predict the quality of software before it is shipped. One of the common approaches to the problem has been to use statistical and machine learning models~\cite{fenton:1999:TSE} with various program features (or quality metrics, e.g., lines of code) and representations (e.g., AST N-grams~\cite{Shippey:2019:IST}). While {\system} does not attempt to address the problem of software defect prediction, the presence of anomalous patterns may be used as yet another program feature to predict software quality.

Specific efforts have also been devoted to the problem of \emph{automatically correcting syntax errors in programs}~\cite{bhatia:2016:arxiv, Yasunaga:2020:ICML}. SynFix~\cite{bhatia:2016:arxiv}, in particular, corrects syntax errors in introductory programming problems by training recurrent neural networks (RNNs) to learn syntactically-valid programs as token sequences. Test programs are then queried against the learnt RNN models to detect syntax errors and predict possible corrections to fix those errors. It looks as if though that the learnt model in SynFix problem-specific: syntax errors in a program meant for some problem can only be fixed by the model learnt from syntactically-valid programs solving the same problem. The decision tree used in {\system} is not problem-specific. DrRepair~\cite{Yasunaga:2020:ICML}, on the other hand, develops an unsupervised learning approach of introducing syntax errors in valid programs and using diagnostic feedback from compilers on those programs to correct them.

{\system} is different than aforementioned approaches in that it is not specific to detecting bugs. An anomaly flagged by {\system} may or may not be a bug --- this largely depends upon the accepted idiosyncratic patterns within a given program's source code. In this sense, {\system} can notify programmers of anomalies, even before test cases or program specifications are checked. To our knowledge, {\system} may be the first of its kind to identify typographical anomalies, which may be erroneous, based entirely on a self-supervised learning system.


\section{Conclusion}

In this study, we presented {\system}, a system to automatically detect possible typographical errors in the control structures of high-level programming languages. {\system} also suggests possible corrections to such errors. The fundamental approach \system\ takes is to recast typographical errors as anomalies, where a self-supervised system that trained on a large enough semi-trusted code will automatically learn which idiosyncratic patterns are acceptable (and which are not). Our findings from scanning C/C++ programs from several open-source packages across 2.57 million programs reveal interesting anomalies (as well as some unusual programming styles). We believe that even when flagged anomalies are not bugs, they may still improve software robustness as was demonstrated by the flagged anomaly in CURL and acknowledged by the developers.

\section{Broader Impact}

{\system} is a machine-learning based self-supervised idiosyncratic pattern detection system that applies learned patterns to detect anomalies in program code. As {\system} uses vast amount of open-source code to learn idiosyncratic patterns, it uses GitHub stars as an indirect measure of quality of a repository. In other words, {\system} treats program code obtained from the open-source repositories as trustworthy. We, nonetheless, believe that {\system} could be susceptible to attacks arising from low-quality and/or malicious data. Specifically, if an attacker is able to control source-code repositories used for training, then {\system} can be easily be fooled to learn otherwise anomalous patterns, which would then flag otherwise non-anomalous patterns as anomalies. Furthermore, as we use the number of occurrences of a pattern to determine anomalies, the attacker does not even need to control multiple repositories, but rather controlling just one repository that has abnormally high occurrences of a malicious pattern would suffice. On a similar note, {\system} could be susceptible to collusion attack in which multiple repositories (attackers) collude to corrupt training data by adding anomalous patterns. We, nevertheless, would like to mention that {\system} is unbiased towards source-code repositories --- all the repositories that meet the criteria of GitHub stars are considered for training.

As {\system} is a machine-learning based system, its output is easy to debug, analyze and explain. Although, its computational demands should be conceptually lower than its deep-learning based version, it, nonetheless, demands increasing computational resources with the growing size of the training dataset. Increasing computational resources can have adverse environmental effects and lead to climate-related issues such as global warming. We, however, believe that this is applicable to any software system that analyzes or learns from data.

\appendix
\section{Appendix: Flagged Anomalies and Possible Corrections}
\label{section:appendix_result}

Below we show some of the interesting anomalies found while scanning OpenSSL and CURL packages.

\noindent\rule{0.47\textwidth}{1pt}\\
\begin{footnotesize}
Potential anomaly: \texttt{((s1 == NULL) $^\wedge$ (s2 == NULL))} \\
Location: \texttt{openssl-1.1.1h/test/testutil/tests.c:268} \\
Possible corrections: \\
\texttt{((s1 == NULL) $^\wedge$ (s2 == NULL))}, edit distance 0, occurrences 8 \\
\texttt{((s1 == NULL) | (s2 == NULL))}, edit distance 1, occurrences 32 \\ 
\texttt{((s1 == NULL) || (s2 == NULL))}, edit distance 2, occurrences 6808 \\
\texttt{((s1 == NULL) \&\& (s2 == NULL))}, edit distance 2, occurrences 521 \\
\end{footnotesize}
\noindent\rule{0.47\textwidth}{1pt}

\begin{footnotesize}
Potential anomaly: \texttt{(-2 == rv)} \\
Location: \texttt{openssl-1.1.1h/test/evp\_test.c:1898} \\
Possible corrections: \\
\texttt{(-2 == rv)}, edit distance 0, occurrences 16529 \\
\texttt{(variable == rv)}, edit distance 1, occurrences 1164852 \\
\texttt{(-2 != rv)}, edit distance 1, occurrences 6483 \\
\texttt{(-2 <= rv)}, edit distance 1, occurrences 2170 \\
\texttt{(-2 >= rv)}, edit distance 1, occurrences 265 \\
\end{footnotesize}
\noindent\rule{0.47\textwidth}{1pt}

\begin{footnotesize}
Potential anomaly: \texttt{(BIO\_puts(bp, ":") <= 0)} \\
Location: \texttt{openssl-1.1.1h/crypto/x509/t\_req.c:141} \\
Possible corrections: \\
\texttt{(BIO\_puts(bp, ":") <= 0)}, edit distance 0, occurrences 475 \\
\texttt{(BIO\_puts(bp, ":") == 0)}, edit distance 1, occurrences 80350\\
\texttt{(BIO\_puts(bp, ":") != 0)}, edit distance 1, occurrences 4559\\
\texttt{(BIO\_puts(bp, ":") < 0)}, edit distance 1, occurrences 1431\\
\end{footnotesize}
\noindent\rule{0.47\textwidth}{1pt}

\begin{footnotesize}
Potential anomaly: \texttt{(s->keepon > TRUE)} \\
Location: \texttt{curl/lib/http\_proxy.c:359} \\
Possible corrections: \\
\texttt{(s->keepon > TRUE)}, edit distance 0, occurrences 4\\
\texttt{(s->keepon > number)}, edit distance 2, occurrences 127540\\
\texttt{(s->keepon > variable)}, edit distance 2, occurrences 56475\\
\end{footnotesize}
\noindent\rule{0.47\textwidth}{1pt}
\section{Appendix: Approaches for Automatic Correction}
\label{section:appendix_autocorrection}

In this section, we provide a brief and informal description of all three approaches that we evaluated to automatically suggest possible corrections to an erroneous pattern. We do not provide a formal description as these approaches are not the contribution of this paper.

For the sake of comparing these approaches, let us consider that the parameters for an automatic correction algorithm consist of (1) target string of length $N$ and its correction of length $M$, (2) a dictionary consisting of $D$ strings among which to search for possible corrections that are within the edit distance of $E$, and (3) the target string and its corrections draw characters from a vocabulary set of size $V$.


    \paragraph{A naive approach.} A naive approach to look for corrections of a target string against a dictionary would be to calculate the edit distance between the target string and every string from the dictionary. 
    
    The time complexity of this approach is linear to the size of the dictionary, and more precisely, it is $O$($N \times M \times D$), where $M$ is the average size of the strings from $D$.
    
    \paragraph{Norvig et al. approach.} An alternative approach suggested by Norvig et al.~\cite{norvig_autocorrection} eliminates the need to go over all the strings from the dictionary to find possible corrections. Instead, it relies on generating candidate correction strings within the given maximum edit distance from the target string. The candidate correction strings are then checked against the dictionary. If a candidate is found in the dictionary, then it is a possible correction of the target string.
    
    The time complexity of Norvig et al. algorithm, however, grows exponentially in the order of the edit distance. Specifically, the number of candidate correction strings that are at edit distance of 1 from the target string are $O$($V \times N$), considering typical typing corrections such as insertion, deletion and replacement of a single character. In order to calculate the candidate correction strings at edit distance of 2, all of the candidate strings at edit distance 1 go through corrections of a single character. In other words, the number of candidate corrections at edit distance of 2 would be $O$($V \times N^2$). This approach works in practice when $N$ is small (for English language, average value of $N$ is 5) and hence $E$ is at max $N$ (typically, in practice, $E$ is 2 or 3 for English language). We found that with a vocabulary size of 50 and the target string of length 80, the algorithm generates ~8000 candidates at edit distance 1 and ~2M candidates at edit distance 2, out of which less than 5\% would be valid candidates.
    
    \paragraph{Symmetric Delete approach.} Symmetric Delete~\cite{symmetric_delete}, introduced by Garbe et al., is another correction candidate generation approach that uses character deletions as edit operations. Although the number of candidates generated from a target string of length $N$ are still upper-bounded by $O$($N!$), they are independent of the vocabulary size $V$. 
    
    The downside of the symmetric delete approach is that it has to generate correction candidates for all the strings from the dictionary. The correction candidates generated using the dictionary (can be pre-computed) are then compared with the candidates generated using a target string to suggest possible corrections to the target string. In other words, it trades memory to store the correction candidates to reduce the time to find possible corrections. The space required to store the pre-computed correction candidates, however, is $O$($D \times M!$), and it proved prohibitive in our case as we increased the number of source repositories.

\balance
\bibliography{main}
\bibliographystyle{plain}

\end{document}